\documentstyle[aas2pp4]{article}
\slugcomment{}
\lefthead{Leitch et al.}
\righthead{Anomalous Galactic Emission}
\begin{document}
\title{An Anomalous Component of Galactic Emission}

\def\simgt{\lower.5ex\hbox{$\; \buildrel > \over \sim \;$}}
\def\simlt{\lower.5ex\hbox{$\; \buildrel < \over \sim \;$}}

\author{E.M. Leitch, A.C.S. Readhead, T.J. Pearson, S.T. Myers\altaffilmark{1}}
\affil{California Institute of Technology, 105-24\\Pasadena CA 91125}
\altaffiltext{1}{University of Pennsylvania}

\begin{abstract}
We present results from microwave background observations at the Owens Valley Radio Observatory.  These observations, at 14.5 and 32~GHz, are designed to detect intrinsic anisotropy on scales of $7^\prime-22^\prime$.  After point source removal, we detect significant emission with temperature spectral index $\beta \simeq -2$ towards the North Celestial Pole (NCP).  Comparison of our data with the IRAS $100\,\mu$m map of the same fields reveals a strong correlation between this emission and the infrared dust emission.  From the lack of detectable H$\alpha$ emission, we conclude that the signals are consistent either with flat-spectrum synchrotron radiation, or with free-free
emission from $T_e\simgt 10^6~$K gas, probably associated with a large HI feature known as the NCP Loop.  Assuming $\beta = -2.2$, our data indicate a 
conversion ${T_{\it f}/I_{100\mu{\rm m}}} = 7.5\times10^{-2}~\nu_{\rm GHz}^{-2.2}$~K/(MJy/sr).

The detection of such a component suggests that we should be cautious in any assumptions made regarding foregrounds when designing experiments to map the microwave background radiation.

\end{abstract}


\section{Observations}

Since 1993, the Owens Valley Radio Observatory (OVRO) 5.5-m telescope (see \cite{Herbig}) has been
used for extensive observations at 32~GHz of the cosmic microwave background 
(CMB) on angular scales of $7^\prime\!-22^\prime$.  The 
receiver input is switched at 500~Hz between two beams of $7^\prime\!.35$ (FWHM) 
separated by $22^\prime\!.16$.  The OVRO 40-m telescope, under-illuminated at 
14.5~GHz to match the 5.5-m beam, provides a second frequency channel for 
spectral discrimination of foregrounds (see Table 1).  Both receivers detect 
right-circular polarization.

\begin{table}[h]
\centerline {\footnotesize\sc TABLE 1}
\centerline {\footnotesize\sc Parameters for the OVRO 5.5-m and 40-m telescopes}
\begin{center}
\footnotesize
\begin{tabular*}{3.0in}[h]{p{0.9in} p{0.9in} p{0.9in}} \hline\hline
               & 5.5-m                 & 40-m  \nl\hline
Frequency      & 32~GHz                & 14.5~GHz              \\
Bandwidth      & 5.6~GHz               & 3~GHz                 \\
System         & 30 K                  & 48 K                  \\
temperature    &                       &                       \\
Beam efficiency& 0.61                  & 0.70                  \\
Beamwidth      &$7^{\prime}.35$        &$7^{\prime}.35$        \\
(FWHM)         &                       &                       \\
Beamthrow      &$22^{\prime}\!.16$     &$21^{\prime}\!.50$     \\
Main beam      &$5.12\times10^{-6}~$sr &$5.38\times10^{-6}~$sr \\
solid angle    &                       & \\\hline
\end{tabular*}
\label{tab:params}
\end{center}
\end{table}

In  the ``RING5M'' experiment, we observe 36 fields at $\delta \simeq 88^\circ$
 spaced by the $22^\prime\!.16$ beamthrow in a ring around the NCP (see Figure 1).  To minimize variations in differential 
contributions from the ground, each field is observed only within $\pm20$~minutes 
of upper culmination.  In each flux measurement, the telescope moves in 
azimuth to alternate the beams on the main field; each measurement is thus
the difference between the signal from the main field and the average of the 
signals in the two adjacent fields (\cite{Readhead}~1989).  As a result, a 
strong signal in one field produces a negative signal half as strong in each 
of the two flanking fields (see Figure 2).

To estimate the contribution of point sources, the RING5M fields were mapped 
on the VLA at 8~GHz to a sensitivity of 0.25~mJy; a total of 34 sources 
were detected with $S_{\rm 8\,GHz} \simgt 2~$mJy.  Subsequent monthly VLA monitoring 
of these sources at 8 and 15~GHz provided accurate measurements of flux 
densities and spectral indices, enabling us to estimate the flux 
densities at both 14.5~GHz and 32~GHz and to subtract the point-source 
contribution from these data sets.    

We report here only on results relevant to an anomalous foreground we have 
detected.  The implications for CMB observations and a full discussion and 
analysis of our results will be presented in subsequent papers.

\section{Analysis}

Our cumulative observations of the RING5M fields have achieved a $1\sigma$ rms sensitivity of $17\,\mu$K per field at 32~GHz, and $15\,\mu$K per field at 14.5~GHz. 
Signals $\geq200\,\mu$K are seen at both frequencies, and excellent 
reproducibility of these data between the 1994, 1995 and 1996 observing 
seasons indicates that they represent real structure on the sky.

In addition to good agreement between independent data sets at the same 
frequency, a Spearman rank test (\cite{Kendall}), modified to account for 
correlations introduced by the switching (see \S3), finds a correlation 
$r_s = 0.84$ between the two frequencies, with a significance 
$P(r_s \geq 0.84) = 7\times10^{-7}$.  The strength of the observed correlation
 between independent observations on separate telescopes is further evidence 
that the signals are astronomical in origin, and not artifacts of the 
observing procedure.

Since the only common element between the two channels is our
observing strategy, we explored the possibility of systematic contamination 
by observing the RING5M fields at 14.5~GHz for two weeks 
at {\it lower} culmination.  The data obtained showed the same signals, to 
within the observed $\sim~\!\!\!80~\mu$K scatter between two-week subsets of our 
upper culmination data.

In all of the following discussion, we use data sets from which the 
point-source contributions have been subtracted.  Apart from one source, which
affected 3 of our 36 fields, the contributions of point sources were much 
smaller than the detected signals.  The maximum $1\sigma$ error on the estimated point-source contribution to a field was $22~\mu$K.  There is therefore no doubt that we have detected a significant astronomical signal which is not due to point sources.

\subsection{Spectral Index of the Foreground}

The strongest signals seen in both the 14.5 and 32~GHz channels have 
steep spectral indices and amplitudes $T\sim300~\mu$K at 14.5~GHz.  A 
likelihood analysis yields $\beta = -1.1^{+0.2}_{-0.3}$ for the spectral index of the data set as a 
whole, indicating the presence of significant emission with $\beta < 0$.  

Recent Westerbork observations (\cite{Wieringa}) reveal polarized structures 
at high Galactic latitude as bright as 8~K at 325~MHz.  These features, on
 scales of $4^\prime-30^\prime$, are seen in linear polarization only; the
corresponding total intensity maps are extremely smooth, and upper limits 
$< 0.5-1~$K are set on any conterparts in total intensity.  This is interesting, 
as $\sim\!8$~K features with spectral index $\beta=-2.7$ can just reproduce the
observed structures at 14.5~GHz if we were 100\% sensitive to linear polarization. 
Tests of our 14.5~GHz polarizers, however, indicate $< 6\%$  
contamination from linear polarization across our bandpass, so it is clear 
that such polarized features cannot account for the signals we have detected.  

Moreover, given the smoothness of the total intensity maps, it is highly improbable that the structure in the polarized emission is due to variations in 
intrinsic polarization angle, and the polarized structure is interpreted as 
Faraday rotation of an intrinsically smooth, polarized synchrotron 
background by an intervening screen.  As a result, the polarization angle 
will have the $\nu^{-2}$ dependence of Faraday rotation and the structure
should be negligible at 14.5~GHz.  

Total intensity maps from the WENSS survey (\cite{de Bruyn}), covering 20 of 
the 36 RING5M fields, show no detectable signals after removal of discrete
sources.  Comparison of the maximum signal at 14.5~GHz with the rms from the 
WENSS maps in the overlap fields places a lower limit $\beta \geq -2.2$ on the spectral index of the foreground we have detected.  

\begin{figure}[ht]
\plotone{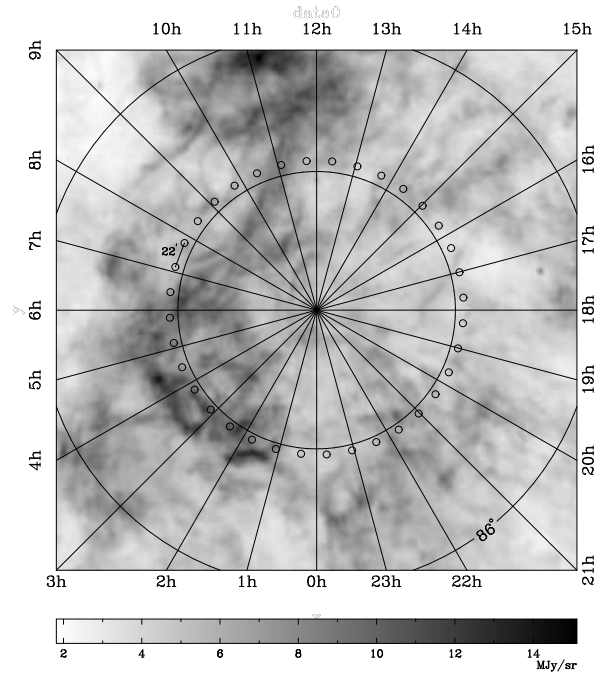}
\caption{\footnotesize IRAS $100\,\mu$m map in J2000 coordinates, with RING5M fields over-plotted.  The spacing of the 
fields is $\sim 22^\prime$, and the FWHM of the beam is $7^\prime\!.35$.}
\label{fig:irasfields}
\end{figure}

Thus, based on the WENSS maps, we know that the contribution of any 
steep-spectrum ($\beta < -2.2$) component is negligible and we now 
investigate what foreground spectral
index is favored by our data.  We model our data as a Gaussian CMB component in the presence of a single foreground of variable strength but constant spectral index $\beta$.  Defining $\Delta T_i = \delta T_i - {{1\over2}(\delta T_{i+1} + \delta T_{i-1})}$, as described in \S1, for each field $i$, we measure
\begin{equation}
\Delta T_i(\nu) = \Delta T_{i_{c\!m\!b}} + \Delta T_{i_{f\!o\!r\!e}}\nu^\beta.
\end{equation}
Given $\Delta T_i(\nu)$ measured at two frequencies $\nu_1$ and 
$\nu_2$, we can solve for the CMB component in terms of the unknown spectral
index of the foreground:
\begin{equation}
\Delta T_{i_{c\!m\!b}}(\beta) = {{\Delta T_i(\nu_1)\nu_1^{-\beta} - \Delta T_i(\nu_2)\nu_2^{-\beta}}\over{\nu_1^{-\beta} - \nu_2^{-\beta}}}.
\label{eq:sep}
\end{equation}
The likelihood function for the CMB component (see, for example, \cite{Readhead}~1989) is then given by
\begin{equation}
\L(\sigma_{c\!m\!b}, \beta) = \prod_{i=1}^{36} {1\over{\sqrt{2\pi(\epsilon^2_i + \sigma^2_{c\!m\!b})}}}\exp\left[-{\Delta T^2_{i_{c\!m\!b}}(\beta)\over{2(\epsilon^2_i + \sigma^2_{c\!m\!b})}}\right],
\end{equation}
where $\epsilon_i$ is the error in the residual CMB component, and $\sigma^2_{c\!m\!b}$ is the intrinsic CMB variance.
The likelihood constructed from the point-source subtracted data sets at 14.5 and 32~GHz peaks at $\beta = -2.25$, with $\beta > -1.8$ ruled out at the 
$1\sigma$ level.  Our data is thus consistent with a foreground of spectral
index $\beta\sim-2$, and we conclude that the foreground is either unusually
flat-spectrum synchrotron radiation, or free-free emission.

\section{IRAS 100 $\mu$m maps}
In an attempt to correlate this component with known Galactic
foregrounds, we convolved the IRAS $100\,\mu$m map (\cite{IRAS}) of the NCP 
with our beam and beam-switch.  The source-subtracted 14.5~GHz data, 
along with results of the IRAS convolution, are shown in Figure 2.  

\begin{figure}[ht]
\plotone{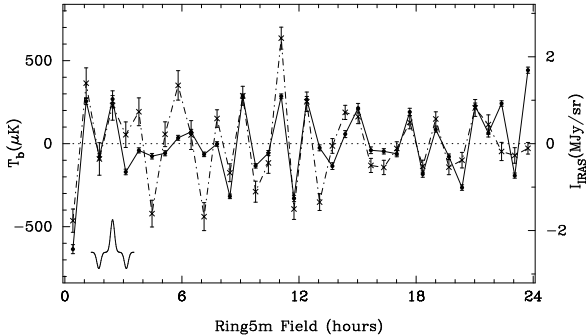}
\caption{\footnotesize Comparison of the 14.5~GHz data (solid line) in $\mu$K, with the IRAS $100\,\mu$m convolution (dot-dashed line).  Errors for the IRAS data points are the estimated standard deviation of the convolution.  The dotted line essentially coincident with the x-axis is the anisotropy inferred from H$\alpha$ images of the NCP fields, in $\mu$K.  At bottom left is the ``triple beam'' pattern due to the double switching.}
\label{fig:kuiras}
\end{figure}

We find a clear correlation between the IRAS $100\,\mu$m maps and our 14.5~GHz
data set.  To assess the significance of this result without {\it a priori}
knowledge of the distribution of the IRAS $100\,\mu$m brightness or 14.5~GHz 
temperature on $7^\prime$ scales, we use Spearman's rank correlation
coefficient $r_s$.  Since this depends only on the data {\it ranks}, whose distribution is known, and not on the values themselves, 
we can determine the significance of an observed value of $r_s$ 
unambiguously.  The observed correlation between the 14.5~GHz and IRAS $100~\mu$m data is $r_s = 0.73$, and, for 36 {\it independent} fields, the probability of observing $r_s \geq 0.73$ by chance is $P(r_s\ge0.73) = 4.5\times10^{-7}$.  
Due to our switching strategy, however, only every third field is actually 
independent, and numerical simulations show that this reduces the significance to $6.7\times10^{-5}$.  

We note that region spanning 3h-8h, where the correlation is weakest, is 
also the region where we see the strongest signals at 32~GHz; the spectral 
indices of these fields are consistent with $\beta = 0$, indicating the 
presence of a significant CMB signal.

Taking $\beta = -2.2$ as the spectral index of the foreground most consistent 
with both our data and the WENSS maps, we can use the 14.5~GHz and 32~GHz data to solve for the free-free component in the manner of Equation (\ref{eq:sep}).  A linear fit to the foreground component yields the conversion
\begin{equation}
{T_{\it f}/I_{100\mu{\rm m}}} = 7.5\times10^{-2}~\nu_{\rm GHz}^{-2.2}~{\rm K/(MJy/sr)}.
\end{equation}

\section{H$\alpha$ Observations of the NCP}

\cite{Gaustad}~(1995) have recently estimated the free-free contamination 
of small-scale anisotropy experiments through H$\alpha$ observations of the 
NCP.  For the brightness temperature of optically thin hydrogen, we expect
\begin{equation}
T_b(\mu{\rm K}) = {5.43\over{\nu^2_{10}T^{1/2}_4}}g_{\it ff}E\!M_{\rm cm^{-6}pc},
\label{eq:tb}
\end{equation}
where
\begin{equation}
g_{\it ff} = 4.69(1 + 0.176\ln{T_4} - 0.118\ln{\nu_{10}})
\end{equation}
is the free-free Gaunt factor (\cite{Spitzer}), the frequency is $10^{10}\nu_{10}~{\rm GHz}$, the electron temperature $T_e = 10^4T_4~{\rm K}$, and $E\!M \equiv \int{n^2_e\,dl}$ is the emission measure.  For $T_e\leq 2.6\times10^4~$K, the H$\alpha$ surface brightness in rayleighs ($1 {\rm R} = 2.42\times10^{-7}~{\rm ergs~cm^{-2}~s^{-1}~sr^{-1}~at~H}\alpha$) is given by
\begin{equation}
I_{H\!\alpha}({\rm R}) = 0.36\,E\!M_{\rm cm^{-6}pc}\,T^{-0.9}_4\label{eq:Ialpha1}
\end{equation} 
(\cite{Shri}).

Simulating our observing procedure on the maps of \cite{Gaustad}~(1995), we measure $\langle\Delta I\rangle_{_{r\!m\!s}}\leq0.1$~R on $7^\prime$ scales in H$\alpha$.  If we assume $T_e = 10^4$~K for the temperature of the emitting gas, the inferred upper limit on the rms at 14.5~GHz due to free-free emission is $\langle\Delta T_{\it ff}\rangle_{_{r\!m\!s}}\leq3.2~\mu$K, a factor of $\sim60$ lower than the observed $\langle\Delta T_{\it ff}\rangle_{_{r\!m\!s}}=203~\mu$K.  
  Furthermore, the H$\alpha$ maps are featureless; in the 36 RING fields, no signals are seen with $|\Delta I| > 0.2~$R.  For the $\sim300~\mu$K signals we detect, Equations (\ref{eq:tb})-(\ref{eq:Ialpha1}) predict an H$\alpha$ brightness $|\Delta I|\sim9~$R.

If considerable dust lies along the line of sight to the NCP, extinction might
account for the low levels of observed H$\alpha$ emission; estimates from 
the IRAS $100\,\mu$m intensities, however, imply $\simlt 0.6$ magnitudes of 
visual extinction (\cite{Simonetti}~1996), so that the upper limits on free-free emission can be increased by 74\% at most.

As $T_e$ is increased beyond $2.6\times10^4~$K, the allowed orbital space for recombination shrinks, and Equation (\ref{eq:Ialpha1}) is no longer valid; for $T_e > 2.6\times10^4~$K, a fit to the H$\alpha$ recombination coefficient gives $\alpha_{H\!\alpha} \propto T^{-1.2}$ (\cite{Ferland}).  The presence of $300\,\mu$K free-free emission can therefore be reconciled
with the observed $3\sigma$ H$\alpha$ limit if the emission is due to gas at $T_e\simgt10^6~$K.  For these temperatures, free-free brightness temperatures of $300\,\mu$K at 14.5~GHz require an $E\!M\simgt 131$.  The corresponding
allowed $n_e-l$ parameter space is shown in Figure 3.  

\begin{figure}[ht]
\plotone{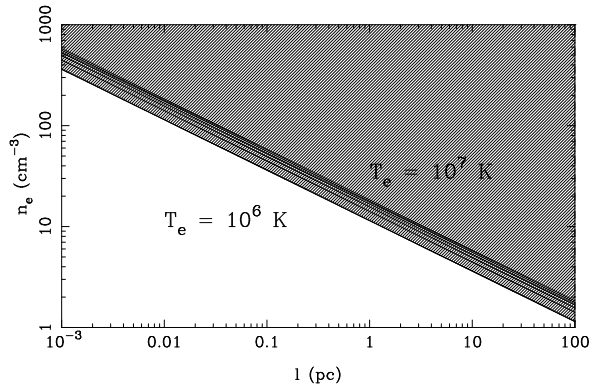}
\caption{\footnotesize Allowed $n_e-l$ parameter space (shaded region) for $T_e\sim10^6-10^7~$K.  We can exclude $l > 100$~pc as this would require that 
inhomogeneities be aligned along the line of sight to within $\sim6\times10^{-3}$~rad, since we see fluctuations on the scale of the $22^\prime$ beamthrow.  The solid lines correspond to $T_e = 10^6,~2\times10^6,~4\times10^6,~6\times10^6,~8\times10^6 {\rm~and~} 10^7~\rm K.$}
\label{fig:nel}
\end{figure}

\section{Discussion}

The observed structure in the IRAS $100~\mu$m map of the NCP 
region is part of a large HI feature known as the NCP Loop.  This feature, 
which encompasses all of the 36 RING5M fields, has been modeled by Meyerdierks
 et al.~(1991) as the wall of an expanding cylindrical shock.  While the 
production of a {\it dense} ionized component such as that implied by Figure 3
 may pose significant difficulties --- such structures will be extremely overpressured, and must necessarily be transient phenomena --- it is intriguing that the combination of large emission measure and high temperature arrived at by 
interpreting the observed structure at 14.5~GHz as free-free emission are 
suggestive of just such a shocked component of the ISM.  

If the emission is due to $\simgt 10^6$~K gas, this component should have a
counterpart in soft X-rays; the absence of any ROSAT PSPC pointings near the 
NCP and the low resolution of available all-sky surveys, however, prevent any 
useful comparison with existing data sets.

Perhaps more plausible, though no less anomalous, is the possibility that the
observed structure at 14.5~GHz is due to flat-spectrum synchrotron emission.  
Synchrotron spectral indices as flat as $\beta = -2.0$ are typically
observed only in plerions associated with the very youngest SNR (\cite{Green}),
and we would not expect such emission from the NCP Loop --- an old remnant, 
with expansion velocity $v\simeq 20$~km/s.  A notable exception to the general
 steepness of Galactic synchrotron radiation, however, is the filamentary structure 
observed toward the Galactic center.  These features, consisting of long, 
nearly one-dimensional threads, have spectral
indices $-2.2 \leq\beta\leq -1.9$ yet show considerable linear polarization,
suggesting that the dominant emission mechanism is synchrotron (\cite{Yusef}).
Although such structures would suffer from a similar lifetime problem as
free-free filaments and require recent injection of high-energy electrons to
maintain a flat spectrum, they would obviate the high temperature and pressure
required in the case of free-free emission.

\section{Conclusions}

We have detected a significant correlation between emission with temperature
spectral index $\beta\sim-2$ observed at 14.5~GHz in the RING5M experiment, and the IRAS $100\,\mu$m maps.  If this is free-free emission, the lack of 
accompanying H$\alpha$ emission implies that it is from a component of the 
ISM with $T_e\simgt 10^6~$K.  The large EM required to 
produce the observed signals at these temperatures is typical of SNR, an 
interpretation supported by the morphology of the NCP Loop, with which the IRAS emission is associated.

\cite{kogut}~(1996) have recently reported a large angular scale correlation of the
residual free-free component in the COBE DMR sky maps with far-infrared DIRBE emission.  The level of this signal at 53~GHz, however, is consistent with 
predictions from H$\alpha$ observations, implying that on $7^\circ$ scales, 
the observed free-free emission is from a $T_e\sim10^4~$K phase of the 
ISM.  Moreover, if the correlation with free-free emission persists to
small scales, the power spectrum of the high-latitude DIRBE $240\,\mu$m maps 
$P({\ell})\propto{\ell}^{-3}$ where ${\ell}\sim 60/\theta^{\circ}$, implies 
a level of free-free at $0.1^{\circ}$ scales marginally consistent with 
the limit inferred from H$\alpha$ observations.

If the observed foreground is not unique to the NCP region, our 
results imply that such emission could be a serious contaminant to 
small-scale CMB measurements in other areas of sky.  This component does, 
however, have a significantly steeper spectral index and may be subtracted out by 
multi-frequency observations.  Moreover, further observations now in progress 
to determine the extent of the correlation between 14.5~GHz and $100~\mu$m 
emission indicate that these results for the NCP region are atypical.

\vskip 12pt

We would like to thank A. G. de Bruyn for making data from the WENSS survey
available to us prior to publication, Gaustad et al. for placing their H$\alpha$ images in the public domain, C. Heiles for pointing us to the work of
Meyerdierks et al, and a referee for a number of useful comments.  This 
research has made use of the SkyView database, operated under the auspices 
of the High Energy Astrophysics Science Archive Research Center (HEASARC) 
at the GSFC Laboratory for High Energy Astrophysics. 
This work at the OVRO is supported by NSF grant number AST 94-19279.

\end{document}